\documentclass[a4paper,final,12pt]{elsarticle}

\usepackage[fleqn]{amsmath}
\usepackage{amssymb}
\usepackage{dcolumn}% Align table columns on decimal point
\usepackage{bm}% bold math
\usepackage{color}
\usepackage{dsfont}
\usepackage{float}%\usepackage[normalem]{ulem}

\usepackage{graphicx,hyperref}% Include figure files
\usepackage{bm}% bold math
\usepackage{subfigure}% bold math

\newcommand \beq{\begin{eqnarray}}
\newcommand \eeq{\end{eqnarray}}
\newcommand \bml{\bar M^2_{\rm L}}
\newcommand \bmt{\bar M^2_{\rm T}}

\newcommand \bgl{\bar G_{\rm L}}
\newcommand \bgt{\bar G_{\rm T}}

\usepackage[normalem]{ulem} % just for strike outs

\newcommand{\rt}[1]{\textcolor{red}{}}

\journal{Nuclear Physics B}

\begin{document}
\allowdisplaybreaks

\begin{frontmatter}



\title{Loss of solution in the symmetry improved $\Phi$-derivable expansion scheme}


\author[a]{Gergely Mark{\'o}}
\ead{smarkovics@hotmail.com}
\author[b]{Urko Reinosa}
\ead{reinosa@cpht.polytechnique.fr}
\author[a]{Zsolt Sz{\'e}p}
\ead{szepzs@achilles.elte.hu}

% The "\note" macro will give a warning: "Ignoring empty anchor..."
% you can safely ignore it.

\address[a]{MTA-ELTE Statistical and Biological Physics Research Group, H-1117 Budapest, Hungary.}
\address[b]{Centre de Physique Th{\'e}orique, Ecole polytechnique, CNRS, Universit{\'e} Paris-Saclay, F-91128 Palaiseau, France.}

\begin{abstract}
We consider the two-loop $\Phi$-derivable approximation for the
$O(2)$-symmetric scalar model, augmented by the symmetry improvement
introduced in Pilaftsis and Teresi (2013)
\cite{Pilaftsis:2013xna}, which enforces Goldstone's theorem in the
broken phase. Although the corresponding equations admit a solution in
the presence of a large enough infrared (IR) regulating scale, we argue that, for smooth ultraviolet (UV) regulators, the solution is lost when
the IR scale becomes small enough. Infrared regular solutions exist for
certain non-analytic UV regulators, but we argue that these solutions
are artifacts which should disappear when the sensitivity to the UV
regulator is removed by a renormalization procedure. The loss of
solution is observed both at zero and at finite temperature, although
it is simpler to identify in the latter case. We also comment on
possible ways to cure this problem.
\end{abstract}

%\pacs{02.60.Cb, 11.10.Gh, 11.10.Wx, 12.38.Cy}                                                 

      

\end{frontmatter}

% \linenumbers


%%%%%
\section{Introduction}

The two-particle-irreducible ($2$PI) effective action formalism and the
corresponding $\Phi$-deriv\-a\-ble expansion scheme are appropriate to
study the dynamical evolution of far from equilibrium quantum systems
\cite{Berges:2015kfa} and some of their equilibrium properties
\cite{Baym:1977qb,Blaizot:2000fc,Berges:2004hn,Blaizot:2005wr}. Despite
their many \mbox{applications}, they suffer, however, from a major problem in
the case of a spontaneously broken continuous symmetry: Goldstone's
theorem is violated by the approximations which casts a doubt on the
obtained physical results. Recently, some approaches have been put
forward to correct for this problem, but they are either restricted to
specific approximations \cite{Ivanov:2005bv,Yukalov:2006wk} or
they require a non-linear representation of the degrees of freedom
\cite{Leupold:2006bp}, which makes the practical implementation
difficult. More recently, a systematic approach based on the usual
linear representation of the fields has been proposed by Pilaftsis and
Teresi \cite{Pilaftsis:2013xna} and applied to various situations
of interest
\cite{Brown:2015xma,Garbrecht:2015cla,Pilaftsis:2015bbs}.

In this paper we discuss this latter approach, the so-called symmetry
improvement, in its application to the O($2$)-symmetric scalar model at
two-loop level, as in \cite{Pilaftsis:2013xna}. We argue that,
although the corresponding equations are formally compatible with
Goldstone's theorem in the absence of an IR regulating scale (``infinite
volume''), they do not have a physical solution (in fact they do not
have a physical solution for a small enough IR scale, that is for a
``large enough volume''), if the UV regulator is chosen smooth enough.
Infinite volume solutions can be found for certain non-analytic UV
regulators and for a fixed value of the cutoff. Even though these
solutions could be of relevance in cases where such UV regulators have
a physical origin, we argue that they should be regarded as artifacts
in applications where the UV regulator has not such origin and that they
should in principle disappear when the dependence with respect to the
UV regulator is eventually removed by a renormalization procedure. We
believe that the absence of infinite volume solutions manifests itself
in other truncations or models when the symmetry improvement is
considered. It illustrates the fact that it may not be enough to
overimpose Goldstone's theorem in order to solve the above mentioned
problem of the $\Phi$-derivable scheme. One might need, in addition,
to make sure that the considered truncation can cope dynamically (and
not just fictitiously through some non-analyticity of the UV regulator)
with the infrared sensitivity that the presence of Goldstone modes
entails.

The implications of the results reported in this paper concerning the
fate of the symmetry improvement should not be considered as definitive,
because we cannot give a complete analytical proof of the loss of
solution. We can only illustrate it using semi-rigorous analytical
arguments, supported by a numerical investigation which, although
compelling, is based on methods that possess their own limitations.
Nevertheless, we believe that in the two-loop approximation considered
in \cite{Pilaftsis:2013xna}, the original formulation of the
symmetry improvement faces problems related to infrared sensitivity. Our
purpose is then to initiate some discussion on this recently proposed
approach that we believe is interesting but calls, in our view, for some
critical discussion. In fact, we have briefly reported on the loss of
solution at finite temperature in Sec.~V.C of
\cite{Marko:2015gpa}.\footnote{More recently, some other unphysical
features of the symmetry improved 2PI approach have been reported, when
it is applied to the study of the linear response of a system
\cite{Brown:2016sak}.} In the present paper, we would like to provide
a more thorough investigation including the $T=0$ case as well.

In Sec.~\ref{s2}, we present the two-loop $\Phi$-derivable equations in the
symmetry improved approach and give a semi-rigorous argument in favor
of the absence of infinite volume solutions. We then support this claim
with a numerical investigation, first at $T=0$ in Sec.~\ref{s3} and then
at finite $T$ in Sec.~\ref{s4}. We finally discuss possible ways to circumvent
the problem.

%s2 #&#
\section{The symmetry improved two-loop $\Phi$-derivable
approximation}\label{s2}

%s2.1 #&#
\subsection{Equations}

In the standard two-loop $\Phi$-derivable approximation of the
$O(2)$-symmetric scalar model, see \emph{e.g.} \cite{Marko:2013lxa},
the so-called gap equations for a fixed value of the field expectation
value $\phi$ (in the presence of a uniform source) are given by
%
%e1 #&#
%e2 #&#
\begin{eqnarray}
\bml(K) & = & m^{2}_{0}+\frac{\lambda_{0}^{(A+2B)}}{12}{\mathcal{T}}[\bgl]
+\frac{\lambda^{(A)}_{0}}{12}{\mathcal{T}}[\bgt]
\nonumber
\\
& + & \phi^{2}\left[ \frac{\lambda_{2}^{(A+2B)}}{12}-\frac{\lambda
^{2}}{72}\big(9{\mathcal B}[\bgl](K)+{\mathcal B}[\bgt](K)\big)\right] ,
\label{eq:bml}\\
\bmt(K) & = & m^{2}_{0}+\frac{\lambda_{0}^{(A)}}{12}{\mathcal{T}}[\bgl
]+\frac{\lambda^{(A+2B)}_{0}}{12}{\mathcal{T}}[\bgt]
%\nonumber\\
 +  \phi^{2}\left[ \frac{\lambda^{(A)}_{2}}{12}-\frac{\lambda
^{2}}{36}{\cal B}[\bgl;\bgt](K)\right] ,\label{eq:bmt}
\end{eqnarray}
where $\bml(K)$ and $\bmt(K)$ are the momentum dependent squared gap
masses, related to the propagators by
%
%e3 #&#
\begin{eqnarray}
\bar{G}_{\text{L,T}}(K)=\frac{1}{K^{2}+\bar{M}^{2}_{\text{L,T}}(K)}\,.
\label{Eq:props_G_LT}
\end{eqnarray}
The bare mass $m_{0}$ and the various bare couplings
$\lambda_{0}^{(A)}$, $\lambda_{0}^{(B)}$, $\lambda_{2}^{(A)}$ and
$\lambda_{2}^{(B)}$ are required for renormalization and $\lambda$ is
the renormalized coupling, see the discussion below and in
\cite{Berges:2005hc,Marko:2012wc,Marko:2013lxa,Patkos:2008ik}. We have
introduced the short-hand notations
%
%e4 #&#
\begin{eqnarray}
\lambda_{i}^{(\alpha A+\beta B)}\equiv\alpha\lambda_{i}^{(A)}+\beta
\lambda_{i}^{(B)}\,,
\end{eqnarray}
with $i=0$ or $2$, as well as
%
%e5 #&#
%e6 #&#
\begin{eqnarray}
{\mathcal{T}}[G] & \equiv& \int_{Q}^{T} G(Q)\,,\\
{\mathcal B}[G_{1}; G_{2}](K) & \equiv& \int_{Q}^{T} G_{1}(Q)
G_{2}(Q+K)\,,
\end{eqnarray}
for the tadpole and bubble integrals, where
%
%e7 #&#
\begin{eqnarray}
\int_{Q}^{T}
f(Q)\equiv T\sum_{n=-\infty}^{\infty}\int\frac{d^{3}q}{(2\pi)^{3}}
f(i\omega_{n},{\bm{q}})\,.
\end{eqnarray}
It is also understood that whenever $\smash{G_{1}=G_{2}}$, we shall
write more simply ${\mathcal B}[G_{1}; G_{1}](K)={\mathcal B}[G_{1}](K)$. We
shall also denote ${\mathcal B}[G_{1};G_{2}](\smash{K=0})$ more simply as
${\mathcal B}[G_{1};G_{2}]$.

The field-dependent propagators $\bar{G}_{\text{L,T}}$ allow us to
construct the effective potential, the extrema of which are given in
terms of the so-called field equation. As we recalled in the
Introduction, a well known problem of the standard $\Phi$-derivable
approach is that, in the broken phase, the transverse propagator
$\bar{G}_{\mathrm{T}}$ evaluated at the minimum $\bar{\phi}$ of the
potential does not fulfill Goldstone's theorem. In the approach
recently proposed by Pilaftsis and Teresi \cite{Pilaftsis:2013xna}, the
field equation (in the broken phase) is replaced by the constraint
$\bar{M}^{2}_{\mathrm{T}}(0)=0$. This constraint, together with
{\eqref{eq:bmt}} can then be rewritten as
%
%e8 #&#
%e9 #&#
\begin{eqnarray}
\bmt(K) & = & -\frac{\lambda^{2}}{36}\bar{\phi}^{2}\Big[{\mathcal B}
[\bgl;\bgt](K)-{\mathcal B}[\bgl;\bgt]\Big]\,,
\label{eq:bmt2}\\
\bar{\phi}^{2} & = & -\frac{\displaystyle
m^{2}_{0}+\frac{\lambda_{0}^{(A)}}{12}{\mathcal{T}}[\bgl]+\frac{\lambda
^{(A+2B)}_{0}}{12}{\mathcal{T}}[\bgt]}{\displaystyle\frac{\lambda
^{(A)}_{2}}{12}-\frac{\lambda^{2}}{36}{\mathcal B}[\bgl;\bgt]}\,
,\label{eq:phi2}
\end{eqnarray}
while the equation {\eqref{eq:bml}} for
$\bar{M}^{2}_{\mathrm{L}}$ remains the same, apart from the change
$\phi^{2}\to\bar{\phi}^{2}$.

Goldstone's theorem seems thus to be obeyed by construction. However
this is only true if the system of equations {(\ref{eq:bml})},
{(\ref{eq:bmt2})} and {(\ref{eq:phi2})} admits a
solution. As we now argue and as we illustrate in the next sections, we
believe that the above equations do not have a solution.

%s2.2 #&#
\subsection{A tentative argument supporting the absence of
solutions}\label{sec:arg}

Let us first mention that systems of equations such as
{(\ref{eq:bml})} and {(\ref{eq:bmt})}, or
{(\ref{eq:bml})}, {(\ref{eq:bmt2})} and
{(\ref{eq:phi2})} are known to have no solution if they are not
properly UV regularized \cite{Reinosa:2011cs}. Thus, in what follows,
we assume that we have chosen some UV regularization of the
sum-integrals $\mathcal{T}$ and~${\mathcal B}$, to which an UV scale or
cutoff $\Lambda$ is associated. For definiteness, we restrict to
regularizations such that each propagator $\bar{G}_{\mathrm{L,T}}(Q)$
that enters an integral is multiplied by a certain regulating function
$R(x)$, with $x=Q/\Lambda $ in the $T=0$ case or $x=q/\Lambda$ in the
$T\neq0$ case,\footnote{The function $R(x)$ should also decrease fast
enough as $x\to\infty$ and be equal to $1$ for $x=0$.} but of course,
for the applications that we have in mind, the final results should not
depend (or depend as less as possible) on the choice of the UV
regulator. We shall come back to this point below.

We shall first consider the case of smooth enough UV regulators. In
this case, our argument will already exclude the existence of solutions
for any given value of the cutoff. We will then see that, for some
other regulators, solutions can exist at a fixed value of the cutoff.
However, we will argue that, in applications where the sensitivity to
the UV regulator has to be reduced by implementing an appropriate
renormalization procedure and by taking large enough values
of~$\Lambda$, these solutions should disappear.

%s2.2.1 #&#
\subsubsection{Smooth enough UV regulators}

Let us first give our argument in the case of smooth enough UV
regulators. Below, we shall be more specific about what we mean by
``smooth enough'', but a typical example is
%
%e10 #&#
\begin{eqnarray}
R_{\rm{smooth}}(x)=\frac{1}{\pi}{\mathrm{Arctan}}\,
\big[{\mathrm{Sinh}}\,[\sigma(1-x)]\big]+\frac{1}{2}\,,
\label{eq:smoothCO}
\end{eqnarray}
with $\sigma$ large enough so that $R_{\rm{smooth}}(0)$ is very close
to $1$.

First, we remark that, because $\bar{G}_{\mathrm{T}}^{-1}(0)$ vanishes
by construction, for Eq.~{(\ref{eq:bml})} to make sense at
$K=0$, the behavior of $\bar{G}_{\mathrm{T}}(K)$ at small $K$ needs to
be anomalous. This is because if it were not anomalous the bubble
contribution ${\mathcal B}[\bar{G}_{\mathrm{T}}](K)$ in this equation
would be infinite for $\smash{K=0}$. The only possibility would then be
that $\phi=0$ but this would mean that the system is in the symmetric
phase at any temperature.

The second step in our argument comes from the remark that an anomalous
behavior for $\bar{G}_{\mathrm{T}}(K)$ is not compatible with
{(\ref{eq:bmt2})} unless 1) the UV regulator generates such
an anomalous behavior or 2) $\bar{M}_{\mathrm{L}}(0)=0$. That the
regulator can generate an anomalous behavior, we shall exclude for the
moment. In fact, this is precisely what we have in mind when we say
that we consider smooth enough UV regulators and
{(\ref{eq:smoothCO})} belongs precisely to this category. As
for $\bar{M}_{\mathrm{L}}(0)=0$, we cannot exclude it a
priori\footnote{We will see later that this does not seem to happen
numerically.} but this would mean that the mass of the longitudinal
mode (which is interpreted as the Higgs particle in some applications
of the O($2$)-model, such as in \cite{Pilaftsis:2013xna}) is always
zero. Thus, in any physically interesting situation where
$\bar{M}_{\mathrm{L}}(0) \neq0$ (and in the presence of a smooth enough
UV regulator), an anomalous dimension for $\bar{G}_{\mathrm{T}}(K)$ is
not compatible with {(\ref{eq:bmt2})}: indeed the bubble
${\mathcal B}[\bar{G}_{\mathrm {L}};\bar{G} _{\mathrm{T}}](K)$
presumably admits a regular expansion at small~$K$, irrespective of the
fact that $\bar{G}_{\mathrm{T}}$ is anomalous or not, because we can
always rewrite the bubble in such a way that the dependence with
respect to the external momentum $K$ is entirely carried by the
infrared regular propagator $\bar{G}_{\mathrm{L}}$.

Based on the above arguments, and as already announced, we expect the
system of equations {(\ref{eq:bml})}, {(\ref{eq:bmt2})}
and {(\ref{eq:phi2})}, not to have a solution if the UV
regulator is smooth enough.

%s2.2.2 #&#
\subsubsection{Other UV regulators}\label{sec:other_reg}

There exist other regulators, however, which allow to circumvent the
previous no-go result, for a fixed value of the cutoff. In some cases,
a non-analyticity of the regulator can generate an anomalous behavior
of the bubble ${\mathcal B}[\bar{G}_{\mathrm{L}};\bar{G}_{\mathrm
{T}}](K)$ at small $K$, even when $\bar{M}_{\mathrm{L}}(0)\neq0$. This
is for instance the case of the regulator
%
%e11 #&#
\begin{eqnarray}
R_{\rm{sharp}}(x)=\Theta(1-x)\,.
\label{eq:sharpCO}
\end{eqnarray}
As we show in \ref{app:sharp}, for such a regulator, already the
zero-temperature perturbative bubble
${\mathcal B}^{T=0}_{\mathrm{pert}}[G;G _{0}](K)$ with
$\smash{G(K)=1/(K^{2}+M^{2})}$ and $ \smash{G_{0}(K)=1/K^{2}}$ is
anomalous,\footnote{We have checked that no such anomalous behavior is
present when the loop integral (and not the propagators forming the
loop) is regularized using a sharp UV regulator.} with
%
%e12 #&#
\begin{eqnarray} \label{eq:anom}
{\mathcal B}^{T=0}_{\mathrm{pert}}[G;G_{0}](K)-
{\mathcal B}^{T=0}_{\mathrm{pert}}[G;G_{0}]\sim-\frac{1}{12\pi^{3}}
\frac{K\Lambda}{\Lambda^{2}+M^{2}}\,,
\end{eqnarray}
as $K\to0$, instead of the normal $\sim K^{2}$ behavior. A~similar
anomalous behavior is obtained at finite temperature, see
\ref{app:sharp}. Moreover, because the anomalous behavior cannot
originate from an anomalous dimension of $\bar{G}_{\mathrm{T}}$ in the
case where $\bar{M}_{\mathrm{L}}(0)\neq0$ (see the argument given
above), we expect the behavior of the bubble
${\mathcal B}[\bar{G}_{\mathrm{L}}, \bar{G}_{\mathrm{T}}](K)$ in the
fully self-consistent case to be given by {(\ref{eq:anom})}
with $M\to\bar{M}_{\mathrm{L}}(0)$.

Then, in this case, the symmetry improved two-loop equations can have
a solution for a fixed value of the cutoff. In particular, for any
application where a UV regulator of the kind discussed here has a
physical origin and the corresponding anomalous term is strong enough,
the symmetry improvement does not suffer from a loss of solution.
However, for the applications that we have in mind in this paper, the
UV regulator has no physical relevance and should be removed eventually,
by ``sending the cutoff to infinity''. In this limit, the prefactor of
the anomalous behavior {(\ref{eq:anom})} approaches zero. In the same
limit, we can also check that the interval over which the anomalous
behavior of the perturbative bubble sets in shrinks. It is thus natural
to expect that above some value of $\Lambda$, the anomalous behavior
is not sufficient to prevent the loss of solution.

It is important to mention that, strictly speaking, the scalar model
that we are discussing possesses a Landau pole, that is a particular
scale $\Lambda_{\mathrm{p}}$ beyond which the cutoff $\Lambda$ should not
be taken.\footnote{It could be that, in the present two-loop
approximation, the Landau pole does not show up in the solution to the
equations, similar to the case seen in \cite{Reinosa:2011cs}, but
it is there in the relation between the bare and the renormalized
coupling(s), see \ref{app:c-terms}.} Therefore, we cannot make
the prefactor of the anomalous behavior as small as we want and it is
difficult to argue in general that the solution disappears above some
value of $\Lambda$. However, in the situation we are focusing on in
this work, the parameters should be such that the scale of the Landau
pole is well separated from the physical scales, to allow for a large
range in the values of $\Lambda$ over which the correlation functions
and the observables are almost insensitive to the UV regulator (after
appropriate renormalization), while not feeling the presence of the
Landau pole, that is ideally one should have $\mu\ll\Lambda\ll
\Lambda_{\mathrm{p}}$. In this case, it is possible to considerably reduce
the prefactor of the anomalous term and we expect then to observe the
loss of solution above some value of $\Lambda$. This would be
consistent with the fact that, in this situation, our conclusions, in
particular the fact that the solution exists or not, should not depend
on the choice of the regulator. After all, the anomalous behavior that
we are discussing here is not generated dynamically (as it would be the
case for a true anomalous dimension), but is a pure regulator effect,
already visible perturbatively. As such, it should disappear, together
with its consequences, in any limit where the results become (almost)
insensitive to the UV regularization. In any case, if the infinite
volume solution would persist at large $\Lambda$, the results would
depend on the chosen UV regulator and the approach would not be
predictive.

%s2.3 #&#
\subsection{Strategy to check the above claims}

Since our arguments above are not based on rigorous mathematical
statements (which are always difficult to construct for non-linear
integral equations), we shall try to check our claims (as much as
possible) by solving the equations numerically. To this purpose we shall
introduce some IR regulator with some associated scale $\kappa$. If
$\kappa$ is large enough, we will typically obtain a solution to the
IR-regularized version of {(\ref{eq:bml})}, {(\ref{eq:bmt2})} and
{(\ref{eq:phi2})}. We will then study how this solution behaves as
$\kappa$ is taken to smaller and smaller values and we will see that
the solution seems to disappear below some non-zero value of~$\kappa$.

Of course, our strategy implicitly assumes that the set of solutions of
the system {(\ref{eq:bml})}, {(\ref{eq:bmt2})} and
{(\ref{eq:phi2})} can be reached from this limiting procedure.
One could imagine a situation where the system at $\kappa=0$ possesses
a solution which cannot be reached by the limit $\kappa\to0$. We cannot
exclude this scenario, but we believe it to be highly improbable, based
on the above argumentation. We also assume that our conclusions should
not depend on the chosen IR-regulator. Since we have no simple way to
show this, we shall work with different types of IR regulators (to be
introduced in the next sections) and compare the corresponding results.
We shall also consider the two types of UV regulators,
{(\ref{eq:smoothCO})} and {(\ref{eq:sharpCO})},
introduced above.

Our expectation at zero temperature, based on the arguments in
Sec.~\ref{sec:arg} is that, as $\kappa$ is decreased, $\bar{M}^{2}
_{\mathrm{L}}(0)$ will follow the logarithmic infrared behavior of
${\mathcal B}[\bar{G}_{\mathrm{T}}](0)$ which, if the coupling is not
large, is given by\footnote{At leading order in the asymptotic
expansion of the perturbative bubble as $\kappa\to0$, $\kappa_{0}$ is
an arbitrary scale whose only purpose is to make the argument of the
logarithm dimensionless. If we would evaluate the next term in the
asymptotic expansion of the perturbative bubble, we could fix
$\kappa_{0}$ to a certain value which depends however on the details of
the regularization of the integral.}
%
%e13 #&#
\begin{eqnarray}
\sim\frac{\lambda^{2}}{1152\pi^{2}}\phi^{2}
\ln\frac{\kappa^{2}}{\kappa^{2}_{0}}\,,
\label{eq:pertIRdep}
\end{eqnarray}
until the solution either disappears in the case of the regulator
{(\ref{eq:smoothCO})} or the anomalous behavior sets
in the case of the regulator {(\ref{eq:sharpCO})},
possibly allowing for a solution in the (``infinite volume'') limit
$\kappa\to0$ at fixed value of the cutoff. In this latter case, this
infinite volume solution should disappear if $\Lambda$ is taken large
enough, in the case where the Landau scale is well separated from the
other scales. In the next section, we shall show the logarithmic
behavior numerically as well as the loss of solution using different
sets of parameters. In Sec.~\ref{s4}, a similar analysis will be done at
finite temperature where the loss of solution is in principle easier to
identify because it is driven by a linearly divergent (instead of
logarithmically divergent) bubble integral.\footnote{We should more
appropriately speak of a linearly (or logarithmically, at $T=0$)
sensitive bubble integral for it never really diverges because the
solution is lost at a finite value of $\kappa$, as we try to
argue.}\looseness=1

%s3 #&#
\section{Loss of solution at zero-temperature}\label{s3}

In this section we consider the system of equations
{(\ref{eq:bml})}, {(\ref{eq:bmt2})} and
{(\ref{eq:phi2})} at zero-temperature with exactly the same
method to remove UV divergences as in \cite{Pilaftsis:2013xna}, as we
recall below, and with exactly the same parameters. We also consider
two other set of parameters where the loss of solution is easier to
identify. As already discussed, it is important to distinguish between
two classes of UV regulators represented by the choices
{(\ref{eq:smoothCO})} and {(\ref{eq:sharpCO})}. For the
smooth regulator, the dimensionless parameter $\sigma$ controlling the
width of the decay of the regulator, is taken equal to $50$. Finally we
shall also play with two distinct IR regularizations. The first one
consists in replacing the constraint $\bar{M}^{2}_{\mathrm{T}}(0)=0$ by
%
%e14 #&#
\begin{eqnarray}
\bar{M}^{2}_{\mathrm{T}}(0)=\kappa^{2}\,,
\label{Eq:IR_reg_mass}
\end{eqnarray}
the second one consists in imposing the constraint as
%
%e15 #&#
\begin{eqnarray}
\bar{M}^{2}_{\mathrm{T}}(|K|=\kappa)=0\,,
\label{Eq:IR_reg_mom}
\end{eqnarray}
and also in regularizing any integral with a minimal lower boundary
$\kappa$.

%s3.1 #&#
\subsection{Renormalization}\label{sec:ren}

We briefly sketch the renormalization of the gap equations
{\eqref{eq:bml}}, {\eqref{eq:bmt}} at arbitrary $\phi$,
using subtractions at zero temperature. Alternatively, one could use
renormalization prescriptions, as in
\cite{Berges:2005hc,Marko:2012wc,Marko:2013lxa}, where these were
imposed at a finite temperature $T_{\star}$ (see also
\cite{Marko:2015gpa}). Some procedures and notations of this latter
approach will be employed also here.

Following the method presented in \cite{Patkos:2008ik}, we start
from the bare gap equations, containing the bare mass and couplings, and
postulate that the explicitly finite equations have exactly the same
form as the bare ones, only that 1) the integrals are replaced by finite
ones, to be determined later and denoted by the index F and 2) that the
bare mass squared and the bare couplings are replaced by the
renormalized mass squared $m^{2}$, which is negative in order to allow
for symmetry breaking, and a unique renormalized coupling $\lambda$.
These finite gap equations are:
%
%e16 #&#
\begin{subequations}
\label{Eq:finite_gap}
%e16.a #&#
%e16.b #&#
\begin{eqnarray}
\bar{M}^{2}_{\text{L}}(K)&=&m^{2}+\frac{\lambda}{4}\big(\phi
^{2}+\mathcal{T}_{\text{F}}[\bar{G}_{\text{L}}]\big)+\frac{\lambda}{12}{\mathcal{T}}_{
\text{F}}[\bar{G}_{\text{T}}]
\nonumber
\\
&-&\frac{\lambda^{2}}{72}\phi^{2}\big(9 {\mathcal B}_{\text{F}}[\bar{G}_{
\text{L}}](K) + {\mathcal B}_{\text{F}}[\bar{G}_{\text{T}}](K)\big)\,,
\label{Eq:finite_gap_L}\\
\bar{M}^{2}_{\text{T}}(K)&=&m^{2}+\frac{\lambda}{12}\big(\phi
^{2}+\mathcal{T}_{
\text{F}}[\bar{G}_{\text{L}}]+3{\mathcal{T}}_{\text{F}}[\bar{G}_{\text
{T}}]\big)
%\nonumber\\
-\frac{\lambda^{2}}{36}\phi^{2}{\mathcal B}_{\text{F}}[\bar{G}_{
\text{L}};\bar{G}_{\text{T}}](K)\,.
\label{Eq:finite_gap_T}
\end{eqnarray}
\end{subequations}

In order to determine the expression of the finite integrals, the
propagators given in Eq.~{\eqref{Eq:props_G_LT}} are expanded around the
auxiliary propagator $G_{\mu}(K)=1/(K^{2}+\mu^{2})$, with $\mu$
playing the role of a renormalization scale, using the formula
%
%e17 #&#
\begin{eqnarray}
\label{Eq:prop_exp}
\bar{G}_{\text{L/T}}(K)=G_{\mu}(K)+(\mu^{2}-\bar{M}^{2}_{\text{L/T}}(K))G_{
\mu}(K)\bar{G}_{\text{L/T}}(K)\,,
\end{eqnarray}
which in the case of the tadpole integral is applied after iterating it
once. After splitting the squared gap masses into local and nonlocal
parts as $\bar{M}^{2}_{\text{L/T}}(K)=\bar{M}^{2}_{\text{L/T,l}}+
\bar{M}^{2}_{\text{L/T,nl}}(K)$, where the nonlocal part contains only
bubble integrals, the expansion {\eqref{Eq:prop_exp}} allows us to
decompose the integrals appearing in the self-energy into divergent and
finite pieces. The divergent pieces are obtained to be:
%
%e18 #&#
\begin{subequations}
\label{Eq:div_pieces}
%e18.a #&#
%e18.b #&#
\begin{align}
&{\mathcal B}_{\text{div}}[\bar{G}_{\text{i}};\bar{G}_{\text{j}}](K)=
{\mathcal B}_{T=0}[G_{\mu}]\,,
\qquad
\label{Eq:div_piece_B}
\\
&\mathcal{T}_{\text{div}}[\bar{G}_{\text{i}}]=\mathcal{T}_{T=0}[G_{\mu}]+(
\mu^{2}-\bar{M}^{2}_{\text{i,l}}){\mathcal B}_{T=0}[G_{\mu}]
\nonumber\\
&
\hspace{1.2cm}+c_{\text{i}}\frac{\lambda^{2}\phi^{2}}{4}
\int_{Q}^{T=0}
G_{\mu}^{2}(Q)\,\Big[{\mathcal B}_{T=0}[G_{\mu}](Q)-{\mathcal B}_{T=0}[G
_{\mu}]\Big]\,,
\label{Eq:div_piece_T}
\end{align}
\end{subequations}
where $\smash{\text{i},\text{j}\in\{\text{T},\text{L}\},}$
$\smash{c_{\text{L}}=1/2,}$ and $\smash{c_{\text{T}}=1/9.}$

At this point, with the help of {\eqref{Eq:div_pieces}} and
without knowing the explicit expression of the counterterms, one can
already write the finite bubble and tadpole integrals appearing in
{\eqref{Eq:finite_gap_L}} and
{\eqref{Eq:finite_gap_T}}. With the notation
$\text{i},\text{j}\in\{\text{T},\text{L}\}$ these read:
%
%e19 #&#
\begin{subequations}
%
%e19.a #&#
%e19.b #&#
\begin{align}
&{\mathcal B}_{\text{F}}[\bar{G}_{\text{i}};\bar{G}_{\text
{j}}](K)={\mathcal B}[
\bar{G}_{\text{i}};\bar{G}_{\text{j}}](K)-{\mathcal B}_{\text{div}}[
\bar{G}_{\text{i}};\bar{G}_{\text{j}}](K)\,,
\\
&\mathcal{T}_{\text{F}}[\bar{G}_{\text{i}}]=\mathcal{T}[\bar{G}_{\text{i}}]-
\mathcal{T}_{\text{div}}[\bar{G}_{\text{i}}]\,.
\label{Eq:finite_Tad}
\end{align}
\end{subequations}
The expression on the right-hand side can be written under a common
integral, as it was done in \cite{Pilaftsis:2013xna}. The procedure
summarized here allows us to explicitly determine the counterterms and
to check that they bring the original gap equations in the postulated
explicitly final form. For completeness, this is done in
\ref{app:c-terms}.

Once the gap equations are renormalized, the value of the field
determined from the constraint $\bar{M}^{2}_{\mathrm{T}}(0)=0$ is given by
the following explicitly finite expression
%
%e20 #&#
\begin{eqnarray}
\bar{\phi}^{2} = -3\frac{12 m^{2} + \lambda\,\big(\mathcal{T}_{\mathrm
{F}}[\bar{G}_{\mathrm L}]+3{\mathcal{T}}_{\mathrm{F}}[\bar{G}_{\mathrm{T}}]
\big)}{3\lambda- \lambda^{2}
{\mathcal B}_{\text{F}}[\bar{G}_{\text{L}};\bar{G}_{\text{T}}]}.
\label{Eq:phi2_F}
\end{eqnarray}
In case of the first IR regularization {\eqref{Eq:IR_reg_mass}}
one has to use on the right hand side $m^{2}\to m^{2}-\kappa^{2}$,
while for the second one
${\mathcal B}_{\text{F}}[\bar{G}_{\text{L}};\bar{G}_{\text{T}}]
\to{\mathcal B}_{\text{F}}[\bar{G}_{\text{L}};\bar{G}_{\text
{T}}](\kappa)$.

%s3.2 #&#
\subsection{Numerical results}

In what follows we present the zero temperature numerical results
supporting our claim that there is no solution to the coupled system of
equations {\eqref{Eq:finite_gap}} and
{\eqref{Eq:phi2_F}}. We first discuss the evidences at the
physical parameters, that is set~A of {Table~\ref{tab:pars}}.
This corresponds to the parameters used in \cite{Pilaftsis:2013xna},
but because we use a different convention, the parameter $\lambda$
appearing there is larger by a factor of 12 than our coupling. Then we
show results for parameters (sets~B and C) which are numerically better
suited to illustrate our point. We mention that if not stated otherwise
the value of the cutoff used is $\Lambda=5~\mbox{TeV}$.

%t1 #&#
\begin{table}
\begin{center}
\begin{minipage}{0.65\textwidth}
\caption{The parameter sets used in the numerical investigation and the
corresponding scale of the Landau pole in units of $\mu
=100~\mbox{GeV}$.}\label{tab:pars}
\begin{center}
\begin{tabular*}{5cm}{llll}
\hline
Set & $\sqrt{-m^{2}}$& $\lambda$ & $\Lambda_{\mathrm{p}}/\mu$ \\
\hline
A & 89 GeV& 1.56 & $\sim10^{66}$ \\
B & 40 GeV& 40.0 & 615 \\
C & 20 GeV& 19.0 & $\sim10^{5}$ \\
\hline
\end{tabular*}
\end{center}
\end{minipage}
\end{center}
\end{table}

%f1 #&#
\begin{figure}
\centerline{\includegraphics[width=0.6\textwidth]{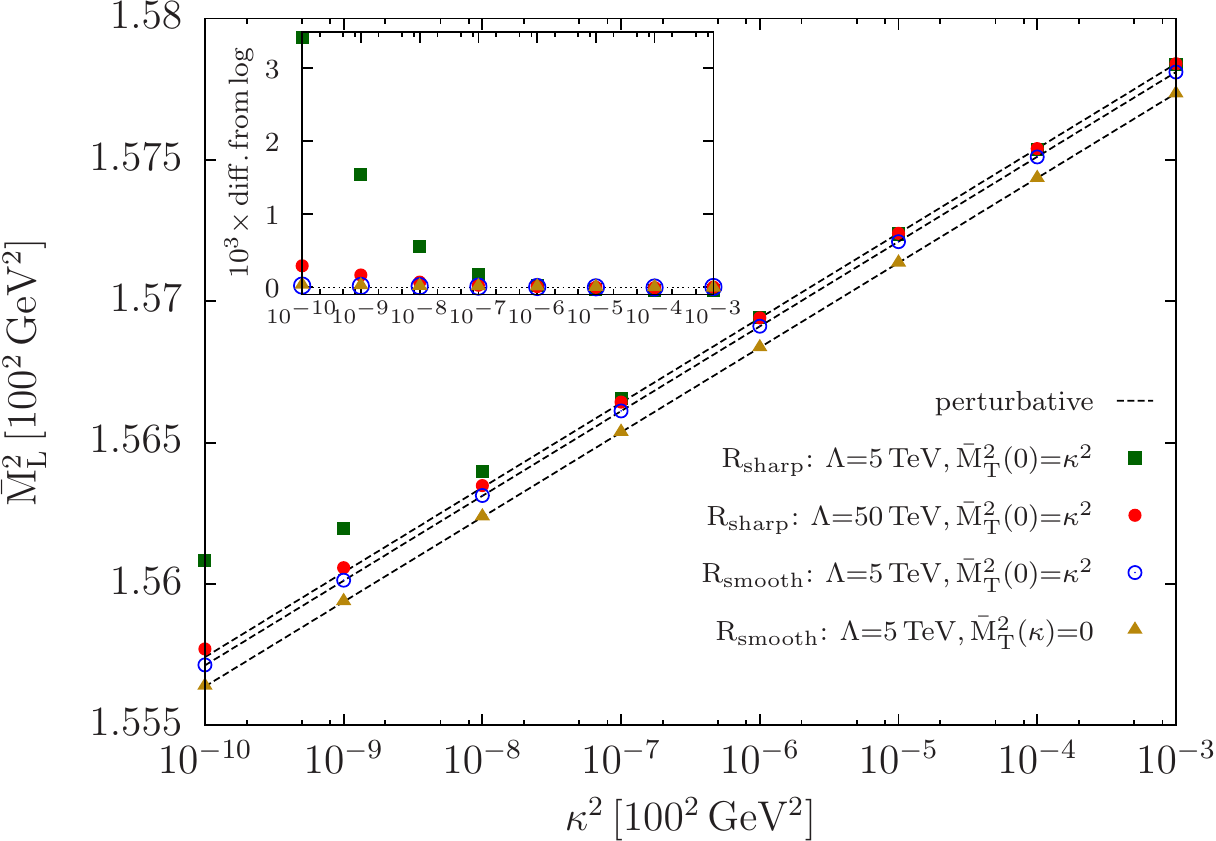}}
\caption{Longitudinal squared gap mass at the lowest available momentum
as a function of the IR regulator parameter. The perturbative
dependence {\eqref{eq:pertIRdep}} is also shown, with
$\kappa_{0}$ fitted. Three different values of $\kappa_{0}$ are needed
for the four sets of points, since only the leading behavior is
regulator-independent. The deviation from the logarithmic behavior of
the points obtained with a sharp UV regulator is caused by the
non-analyticity of the regulator. The inset shows the difference of
each point set to the corresponding perturbative (logarithmic)
behavior.} \label{fig:ML2_of_IRreg}
\end{figure}

%s3.2.1 #&#
\subsubsection{Physical parameters}

In {Fig.~\ref{fig:ML2_of_IRreg}} we compare
$\bar{M}_{\mathrm{L}}^{2}$ at the lowest available momentum bin in
different UV and IR regularizations. Concerning the numerical
implementation of the constraint $\bar{M} ^{2}_{\mathrm{T}}(0)=0$, in
the case of the green squares, red blobs and blue circles we use the IR
regularization {\eqref{Eq:IR_reg_mass}}, whereas for the yellow
triangles we use {\eqref{Eq:IR_reg_mom}}. All four sets of
points are compared to the corresponding perturbative result which
displays a logarithmic dependence with respect to the IR regulator.
This comparison reveals the following. The results obtained using a
sharp UV regularization deviate at some point from the logarithm and,
in the case of the smaller cutoff, they seem to reach a plateau,
suggesting that an IR limit exists, as anticipated above. However,
increasing the value of the cutoff makes the deviation to occur at
smaller $\kappa$, which shows that the effect of the anomalous behavior
is reduced as the cutoff is increased and suggests that above some
value of the cutoff, the behavior of the solution as $\kappa$ is
decreased should be essentially the same as that with a smooth UV
regulator. In this latter case, the results follow the perturbative
logarithmic behavior down to values of~$\kappa$, where we reach the
limits of our numerical precision. This can be seen also in the inset
of {Fig.~\ref{fig:ML2_of_IRreg}} where the difference to the
perturbative result is shown.

%f2 #&#
\begin{figure}
\centerline{\includegraphics[width=0.6\textwidth]{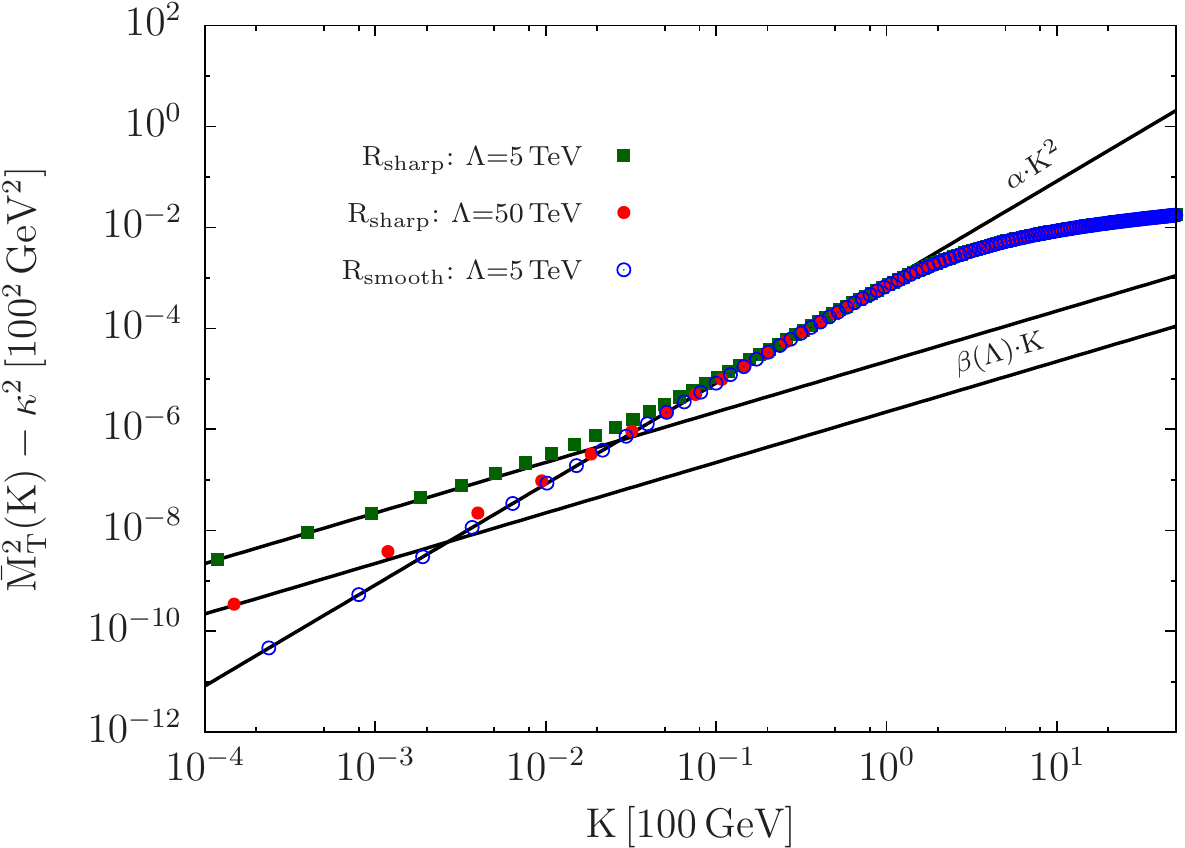}}
\caption{Deep IR behavior of the transverse self-energy. As the ultraviolet
cutoff is increased, the importance of the anomalous, non-quadratic behavior in
the case of a sharp UV regulator is reduced and the results approach those
obtained with the smooth UV regulator. The coefficient $\beta(\Lambda)$ is the
perturbative one which follows from {\eqref{eq:anom}} with
$M^{2}\to\bar{M}^{2}_{\mathrm{L}}(0)$ while $\alpha= 8.27\cdot10^{-4}$ is
obtained through fitting.}
\label{fig:MT2_of_momentum}
\end{figure}

The different behavior displayed by $\bar{M}_{\mathrm{L}}^{2}$,
depending on the chosen UV regulator, can be understood from the IR
momentum behavior of $\bar{M}_{\mathrm{T}}^{2}(K)$, in line with the
discussion of the previous section. This is shown in
{Fig.~\ref{fig:MT2_of_momentum}}, where one can see that in the
case of a sharp UV regulator, a linear term is present in the
transverse squared gap mass at small $K$, with a prefactor which
follows exactly from {(\ref{eq:anom})}. In contrast, with a
smooth UV regulator the transverse self-energy depends quadratically on
the momentum in the deep IR and therefore according to the argument of
Sec.~\ref{sec:arg} we expect a loss of solution in this case.
Unfortunately, for the parameters used in
{Figs.~\ref{fig:ML2_of_IRreg} and \ref{fig:MT2_of_momentum}}
(set~A of {Table~\ref{tab:pars}}) the numerical check of the
loss of solution turned out to be beyond reach, as we are limited by
numerical precision. To visualize the loss of solution, we now switch
to better suited sets of parameters.

%s3.2.2 #&#
\subsubsection{Illustrative parameters}

Using the parameter sets~B and C given in
{Table~\ref{tab:pars}}, we now show that, at $T=0$, there are
two different ways in which the physical solution is lost. We define
$\kappa_{\varnothing}$ as the value of the IR regulator parameter at
which the solution is lost. By losing the solution we mean that, for
$\kappa< \kappa_{\varnothing}$, the coupled gap equations together with
the constraint which determines the field expectation value have no
solution which satisfies $\bar{\phi}^{2}>0$ and that can be reached by
our iterative method. Of course, there could be solutions not reachable
by the iterative method, but we believe that this is not the case for
the physical solution. For the parameter set~B, the loss of the
physical solution happens due to the logarithmic IR sensitivity of the
transverse bubble, while for the parameter set~C, $\bar{\phi}^{2}$
becomes negative at $\kappa_{\varnothing}$.

%f3 #&#
\begin{figure}
\centerline{\includegraphics[width=0.6\textwidth]{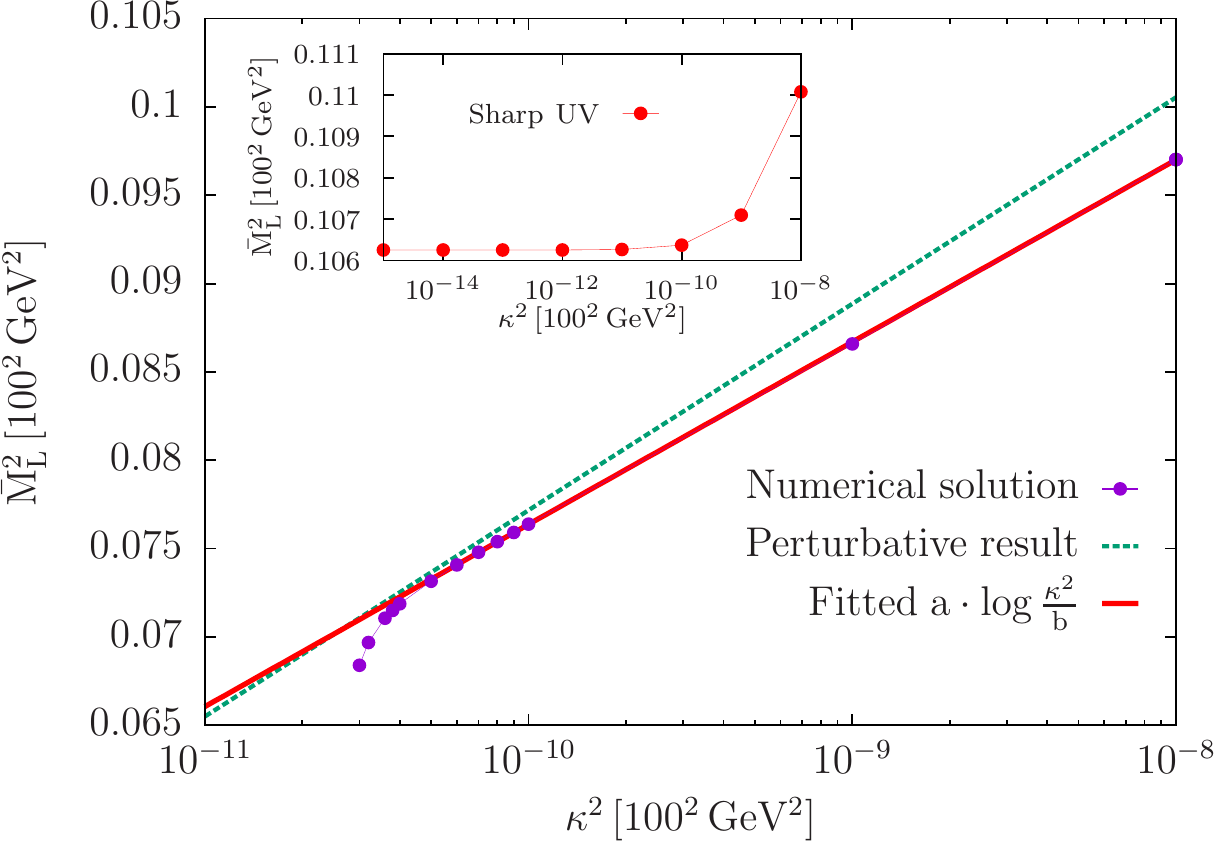}}
\caption{Loss of solution with decreasing $\kappa$ using parameter
set~B. The points are obtained with the IR regularization scheme
{\eqref{Eq:IR_reg_mass}}, using $N_{\mathrm{S,cub}}=501$ and
the smooth UV regulator {\eqref{eq:smoothCO}} with
$\Lambda=5~\mbox{TeV}$. The dashed green line is the perturbative
expression {\eqref{eq:pertIRdep}}, while the red line is the
result of a fit. (For interpretation of the references to
color in this figure legend, the reader is referred
to the web version of this article.)} \label{fig:LosOrig}
\end{figure}

First we discuss the results obtained using parameter set~B. In this
case, the coupling was increased compared to set~A in an attempt to
enhance the logarithmic sensitivity as much as possible, while keeping
the scale of the Landau pole far enough from the physical scales, see
{Table~\ref{tab:pars}}. We expected that at large coupling it
would be easier to capture numerically the effect of the IR sensitivity
than it was the case for the parameter set~A. As shown in
{Fig.~\ref{fig:LosOrig}}, the solution in the case of a smooth
UV regulator is indeed lost below some $\kappa_{\varnothing}$. The
logarithmic behavior of $\bml$ as a function of $\kappa^{2}$ abruptly
changes, just above $\kappa_{\varnothing}$. This points in the
direction that the loss of solution at small $\kappa$ is indeed caused
by the logarithmic IR sensitivity, which in the end ruins the solution
of the equation. Similar features have been observed in the
conventional $\Phi$-derivable expansion scheme, when trying to approach
a critical point \cite{Marko:2015gpa}. Notice that the coefficient
multiplying the logarithm changes, compared to the perturbative one. We
believe that this is due to the non-perturbative corrections the
self-energies receive. In the case of a sharp regulator, with
$\Lambda=5~\mbox{TeV}$, it seems that we have a solution down to
$\kappa=0$, as shown in the inset of
{Fig.~\ref{fig:LosOrig}}.\footnote{There is a few percent
difference between the value of $\bml$ shown in the inset compared to
the value in the main plot which originates from the different UV
regulators used in the two cases. For larger $\kappa$ values this
difference diminishes.} However, the proximity of the Landau pole
($\Lambda_{ \mathrm p}/\Lambda\sim10$) does not allow us to test
whether this solution disappears for large enough~$\Lambda$. In order
to test this scenario, one should choose parameters (smaller values of
$\lambda$) such that $\Lambda_{\mathrm{p}}$ is much further away from
the physical scales. In such cases, the scale $\kappa_{\varnothing}$
becomes smaller and smaller, and it is difficult to see the loss of
solution numerically. But as already explained in
Sec.~\ref{sec:other_reg}, in any case, the sharp UV regulator leads to
inconsistent results.

%f4 #&#
\begin{figure}
\centerline{\includegraphics[width=0.6\textwidth]{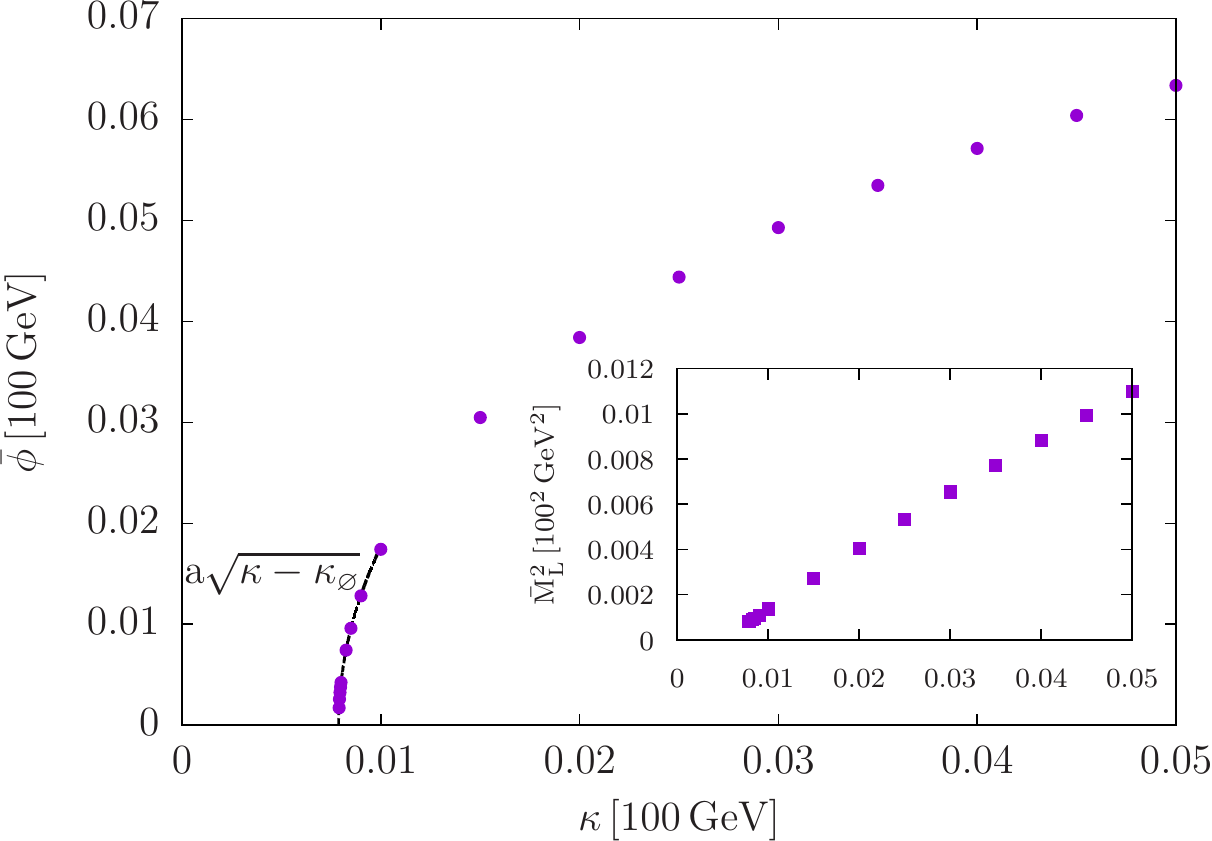}}
\caption{Loss of solution with decreasing $\kappa$ using parameter set C. More
details on the mechanism are given in the text.}
\label{fig:phi2to0}
\end{figure}

Turning now to the results obtained with the parameter set C, we show
in {Fig.~\ref{fig:phi2to0}} the field expectation value as a
function of~$\kappa$. Close to $\kappa_{\varnothing}$,
$\bar{\phi}(\kappa) \sim\sqrt{\kappa-\kappa_{\varnothing}}$ is a good
estimate, as indicated by the fit, which also suggests the
disappearance of the solution. The inset shows that
$\bar{M}_{\mathrm{L}}^{2}$, although decreasing linearly, is still
non-vanishing at $\kappa_{\varnothing}$. The simple reason why this
happens is that for the parameters of set C no broken phase exists. The
criterion for a broken phase to exist at zero temperature (forgetting
about IR problems for a moment) is that $T_{c} > 0$, of course. Since
at the critical temperature $\bar{
\phi}=\bar{M}_{\mathrm{L}}=\bar{M}_{\mathrm{T}}=0$, a very simple
equation determines the line in the parameter space where the system is
critical at zero temperature. This equation is obtained using the gap
equations at $\phi=0$ with vanishing masses. It is found using a sharp
regulator that
%
%e21 #&#
\begin{eqnarray}
-m^{2} = \frac{\lambda}{3}{\mathcal{T}}_{F}[G_{0}] =
\frac{\lambda}{48\pi^{2}}\frac{\Lambda^{2}\mu^{2}}{\Lambda^{2}+\mu^{2}}
\equiv-m_{\mathrm{c}}^{2}(\lambda).
\end{eqnarray}
The result obtained with a smooth regulator with cutoff $\Lambda$ is
practically the same, however it is only available numerically.

For $m^{2}<m^{2}_{\mathrm{c}}(\lambda)$ the critical temperature is
larger than zero, thus it is meaningful to search for a $\bar{\phi}^{2}
> 0$ solution, on the other hand, if
$m^{2}>m^{2}_{\mathrm{c}}(\lambda)$, there is only the symmetric phase
solution ($\bar{\phi}=0$) at $T=0$ with
$\bar{M}_{\mathrm{L}}=\bar{M}_{\mathrm{T}} > 0$. Plugging the
parameters of set C into the expression of
$m^{2}_{\mathrm{c}}(\lambda)$ we see that, indeed, only a symmetric
phase could exist at sufficiently large $\Lambda$.

%s4 #&#
\section{Loss of solution at finite temperature}\label{s4}

In this section we investigate the loss of solution at finite
temperature, numerically. We choose a representative temperature value
$T=30~\mbox{GeV}$ and, using the parameter set~A, we monitor the solution as a
function of~$\kappa$. Due to the employed numerical method, which is
based on Fast Fourier Transform (FFT) (for more details, see
\ref{app:numerics}), we must use a combination of the IR regularization
schemes of {\eqref{Eq:IR_reg_mass}} and
{\eqref{Eq:IR_reg_mom}}, that is
%
%e22 #&#
\begin{eqnarray}
\bmt\left( \omega_{n}=0,\frac{\Lambda}{N_{s}}\right) =\kappa
^{2},\label{Eq:IR_reg_mass_mom}
\end{eqnarray}
where, because the Fourier grid should resolve the smallest available
scale $\kappa$, we should require that the ratio $c=\kappa N_{s}/
\Lambda$ of this scale to the lattice spacing $\Delta k=\Lambda/N
_{s}$ be ideally much greater than $1$. We use a finite number
$N_{\tau}$ of Matsubara modes and we check that the dependence of our
results on $N_{\tau}$ is negligible. Also, in contrast to the $T=0$
case, we apply a 3d UV regulator $R(q/\Lambda)$. In the case of the
sharp regulator, the bubble of {\eqref{Eq:div_piece_B}} reads
%
%e23 #&#
\begin{eqnarray} \label{eq:B}
{\mathcal B}_{T=0}[G_{\mu}]=\frac{1}{8\pi^{2}}\left( \ln\frac
{\Lambda+\varepsilon_{
\Lambda}}{\mu}-\frac{\Lambda}{\varepsilon_{\Lambda}}\right) \,,
\end{eqnarray}
with $\varepsilon_{\Lambda}=\sqrt{\Lambda^{2}+\mu^{2}}$. We also have
to adjust the calculation of the finite tadpole {\eqref{Eq:finite_Tad}}
in the following way. To enhance the $N_{\tau}$ convergence we compute
the difference of tadpoles as
%
%e24 #&#
\begin{eqnarray} \label{eq:dT}
{\mathcal{T}}[\bar{G}_{\text{L/T}}]-\mathcal{T}_{T=0}[G_{\mu}]&=&\int
_{Q}^{T}\big[\bar{G}_{
\text{L/T}}(Q)-G_{\mu}(Q)\big]
%\nonumber\\
+\frac{1}{2\pi^{2}}\int_{0}^{\Lambda}d q\, q^{2}\frac{n_{
\varepsilon_{q}}}{\varepsilon_{q}},
\end{eqnarray}
where $\varepsilon_{q}=\sqrt{q^{2}+\mu^{2}}$ and
$n_{x}=1/(\exp(x/T)-1)$ is the Bose--Einstein statistical factor. For
the second line of {\eqref{Eq:div_piece_T}} we use the relation
%
%e25 #&#
\begin{eqnarray}
\int_{Q}^{T=0}  G_{\mu}^{2}(Q)\,\Big[{\mathcal B}_{T=0}[G_{
\mu}](Q)-{\mathcal B}_{T=0}[G_{\mu}]\Big]
%\nonumber
%\\
\equiv-\left[ \frac{1}{3}\frac{\partial{\mathcal{S}}_{T=0}[G_{\mu
}]}{\partial\mu^{2}}+({\mathcal B}_{T=0}[G_{
\mu}])^{2}\right] ,
\end{eqnarray}
with $\mathcal{S}_{T=0}[G_{\mu}]$ taken from equation (B20) of
\cite{Marko:2012wc}, where it was calculated with a 3d cutoff.

We mention that the calculation of the perturbative bubble done in
\ref{app:sharp} with a 3d sharp UV regulator reveals, in its vacuum
part, a linear dependence with respect to small external momenta (see
{\eqref{eq:finiteT_linearDep}}), similar to the one obtained
with a 4d regulator. The thermal part also displays such a dependence
but with a prefactor which is suppressed by a factor of
$\exp(-\Lambda/T)$. As in the $\smash{T=0}$ case, this has to be
considered an artifact and from now on we switch to a smooth UV
regulator. In this case, we replace the expression in
{(\ref{eq:B})} and {(\ref{eq:dT})} with integrals
calculated using $R_{\text{smooth}}(q/\Lambda)$.

In the limit $c\to\infty$ (and $\kappa$ kept small) the IR
regularization {\eqref{Eq:IR_reg_mass_mom}} results in $\bml$
being dominated by the non-perturbative bubble
${\mathcal B}[\bar{G}_{\mathrm {T}}]$ at zero-momentum, with
$\bmt(0,0)=\kappa^{2}$. Thus, using the high temperature expansion in a
perturbative zero-momentum bubble with mass $\kappa$ gives us the
perturbative estimate of the dependence of $\bml$ for small $\kappa$
(before the loss of solution):
%
%e26 #&#
\begin{eqnarray}
\sim\frac{\lambda^{2}}{576\pi}\phi^{2}\frac{T}{\kappa}+\mathit{const.}\ ,
\label{eq:pertIRdep_atT}
\end{eqnarray}
%
%
%f5 #&#
\begin{figure}
\centerline{\includegraphics[width=0.6\textwidth]{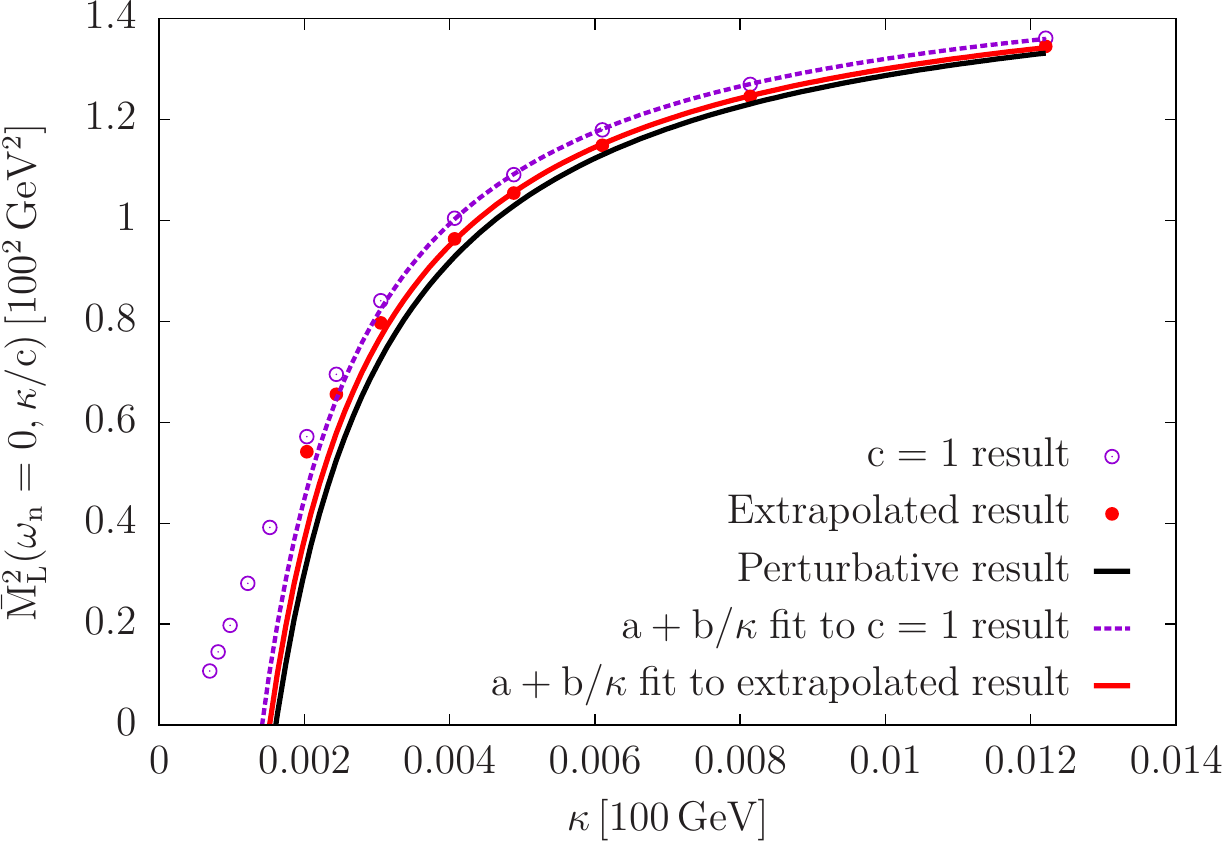}}
\caption{Loss of solution with decreasing $\kappa$ using the parameter set~A at
a temperature $T=30~\mbox{GeV}$. Notice that $\bml$ does not become zero. The
perturbative result is given in {\eqref{eq:pertIRdep_atT}}.
The red blobs come from extrapolating to $c=\infty$ results obtained
at finite $c$. (For interpretation of the references to
color in this figure legend, the reader is referred
to the web version of this article.)}
\label{fig:fTLoss}
\end{figure}%
which foretells a faster loss of solution than in the zero temperature
case due to the stronger (linear) sensitivity of the transverse bubble
diagram. Indeed, as shown in {Fig.~\ref{fig:fTLoss}} the solution is lost
at some $\kappa_{\varnothing}$, where still $\bml\neq0$.\vspace{1pt} In practice
this means that we found no iterative solution with the lowest value
$\alpha=10^{-3}$ used in the under-relaxation method (see Eq.~(135) of
\cite{Marko:2012wc}). It also can be seen that when $a+b/\kappa
$ is fitted to the data, the extrapolation to $c=\infty$ of results
obtained at $c\in\{1,2,4,8\}$ pushes the coefficient of $\kappa^{-1}$
of the fit closer to its perturbative value. Furthermore, we see that
the qualitative behavior of the solution changes before the solution is
truly lost. This is probably due to the inherent inaccuracy of the
discrete sine transform for light modes with small
momenta.\footnote{Using rotational symmetry, the 3d Fourier transform
becomes a sine transform. Its sensitivity can be illustrated in the case
of a massless propagator: one has $\int_{0}^{\Lambda}d p \sin(x p)/p=
\mathrm{Si}(\Lambda x)$, which tends to $\pi/2$ as $\Lambda\to
\infty$ through wild oscillations with slowly decreasing envelope.}

%f6 #&#
\begin{figure}
\centerline{\includegraphics[width=0.6\textwidth]{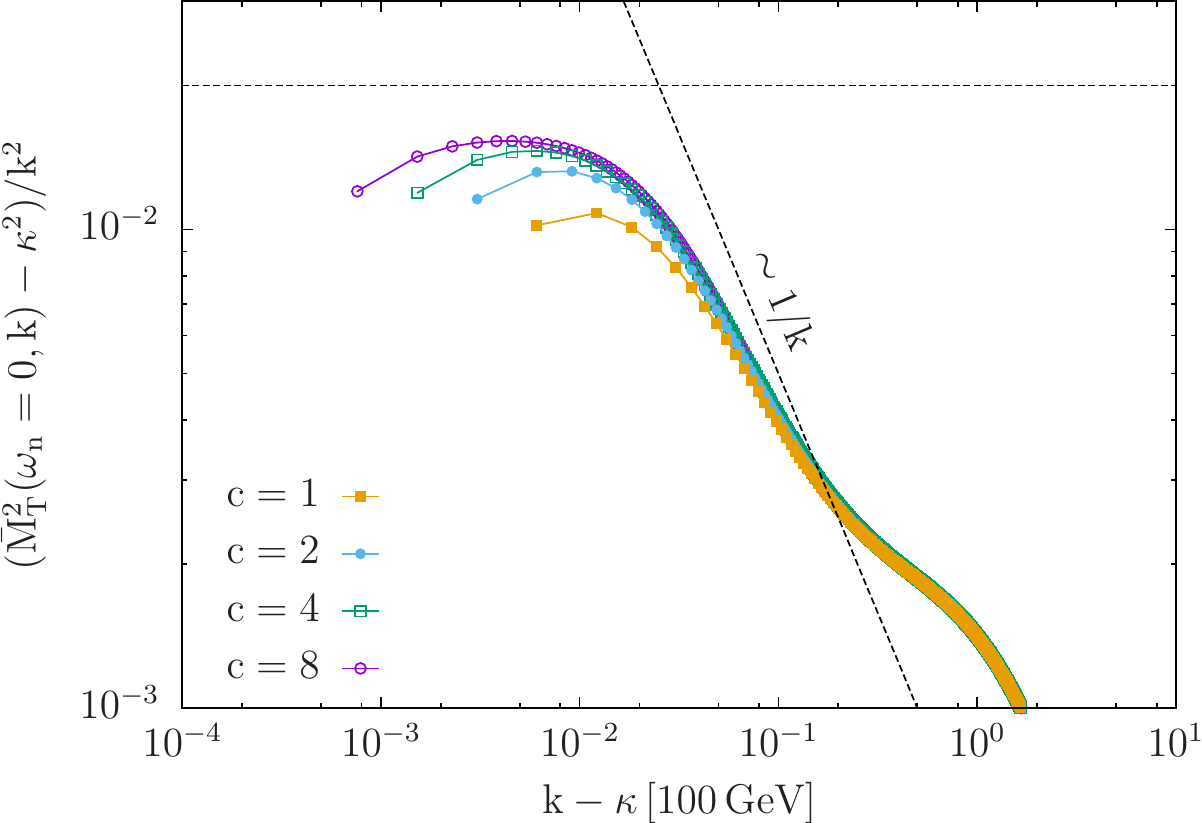}}
\caption{The deep IR behavior of $\bmt$ at the leftmost red blob of
{Fig.~\ref{fig:fTLoss}}, where $\kappa\approx0.002\times
100~\mbox{GeV}$. Notice that the y-axis is scaled by $k^{2}$.}
\label{fig:fT_IR_MT2}
\end{figure}

Another evidence that suggests that an infinite volume limit cannot be
reached is the IR behavior of the self-energy $\bmt(\omega_{n}=0,k)$,
which we show in {Fig.~\ref{fig:fT_IR_MT2}}. There, we see that
the momentum dependence in the deep IR gradually conforms to $\sim
k^{2}$, as $c$ is increased. As far as this behavior persists,
according to our argument in Sec.~\ref{sec:arg}, a loss of solution is
to be expected at small enough $\kappa$. This was observed already in
\cite{Marko:2015gpa} (see Fig.~16 there) with a cruder IR
regularization. There, instead of the regularization
{\eqref{Eq:IR_reg_mass_mom}}, we used
$\bar{M}^{2}_{\mathrm{T}}(\omega_{n}=0, \Delta k)=0$, which enhances
the error of the discrete sine transform, as the lowest momentum mode
becomes massless. %\looseness=1

%s5 #&#
\section{Conclusions}\label{s5}

We studied the implications of the symmetry improved approach, recently
proposed in \cite{Pilaftsis:2013xna}, on the solution of the
$O(2)$-symmetric scalar model, by using a truncation at two-loop level
of the 2PI effective action.

We argued that, due to an untamed IR sensitivity at this approximation order, the constraint imposed
in this approach on the transverse gap mass
(\hbox{$\bar{M}_{\mathrm{T}}(K=0)=0$}) leads to a loss of the solution to
the set of coupled gap equations, unless an anomalous dimension is generated. This seems not
to be the case for smooth enough UV
regulators, where our numerical study, both at zero and finite
temperature, supports our claim that the solution is lost. We tried
various numerical implementations of the above constraint, all involving
an IR regulating parameter $\kappa$, and showed that when one tries to
reach the $\kappa\to0$ limit, the iterative solution is lost at some
finite $\kappa$. At zero temperature, the way this loss of solution
happens in different regions of the parameter space lead us to believe
that this is a general feature of the symmetry improved approach at the
present level of truncation. It has to do with the numerically observed
logarithmic IR sensitivity and is not related to the iterative method
used. At finite temperature, the IR sensitivity is stronger (linear),
however, our FFT based numerical method makes accurate investigations
more difficult. In this case, although the loss of solution is quite
convincing and in line with the preliminary finding of our previous
study \cite{Marko:2015gpa}, we cannot exclude in principle the
possibility that the absence of solution is due to the fact that the
iterative method cannot converge, hence further investigation with a
different numerical method is required. For certain non-analytic UV
regulators, an anomalous dimension is present which, if strong enough,
can lead to an infinite volume solution for a fixed value of the UV
cutoff. We argued that this solution should be considered as an artifact
in any situation where the dependence to the UV regulator can be removed
or considerably reduced by taking large enough values of $\Lambda$
(although it could be relevant in situations where such a UV cutoff has
a physical origin).

We mention that there can be parameters, to which physical parameters
may or may not belong, at which a loss of solution can be observed
numerically only if an extremely fine resolution can be achieved in the
deep IR. In those cases our present investigation should be regarded as
a warning: when a self-consistent propagator equation obtained in a
given truncation is unable to generate an anomalous dimension, then IR
divergences could remain untamed. Therefore, it would be interesting to
test the features of the symmetry improved approach at higher level
truncation of the 2PI effective action, in particular in the
next-to-leading order of the 2PI-$1/N$ expansion, where an anomalous
dimension is known to be generated at the critical point.

Another possible way to circumvent the difficulty is to recognize that
the infinite volume limit is just a convenient mathematical limit (when
it can be taken) but, in principle, any physical system has a finite
size. It may then be that, in some cases of interest, the loss of solution
occurs at a value of $\kappa_{\varnothing}$ way below the inverse
linear size $\pi/L$ of the system. For instance, and without trying to
be rigorous, in the case of the application to Higgs physics of
\cite{Pilaftsis:2013xna}, a rough estimate of the maximal linear size
obtained from $\kappa_{\varnothing}$ using a simplification
(localization as in \cite{Marko:2015gpa}) of the equations is
still more than 200 orders of magnitude larger than the size of the
observable Universe. Keeping in mind that by doing such estimates one
might overpass the validity of this field theoretical model, one can
reverse this argument, which then tells us that using the parameters of
\cite{Pilaftsis:2013xna} and taking $\kappa$ to be inversely
proportional to the scale of the observable Universe would result in a
$\approx5\%$ underestimation of the Higgs mass.

\section*{Acknowledgments}
U. Reinosa would like to thank J. Serreau for discussions on the
spurious effects related to the use of sharp UV regulators. G.M. and
Zs.Sz. were supported by the
Hungarian Scientific Research Fund (OTKA) under Contract
No.~K104292. The work was supported in its initial
stages by the Hungarian--French Collaboration program
T{\'{E}}T\_11-2-2012 (PHC Balaton No.~27850RB).

%sA #&#
\appendix
\section{Numerics}\label{app:numerics}

In this Appendix, the interested reader may find more details on the
numerical procedure we carried out. We concentrate on the $T=0$ case.
For the finite temperature analysis we direct the reader to the
original paper \cite{Marko:2012wc} where various techniques that we
employ here have been introduced and described in
detail. However we point out that we use fast Fourier
transform to compute convolutions, which restricts our choice of
discretization to a uniform grid, with a non-zero first momentum bin.
It also needs to be mentioned that the convergence of the Fourier
transform is only ensured for massive propagators. These two features
dictate our choice of IR regularization scheme
{\eqref{Eq:IR_reg_mass_mom}}.

We either solved the coupled gap equations at fixed $\phi$ (that is
{\eqref{Eq:finite_gap_L}} and
{\eqref{Eq:finite_gap_T}}) or together with
{\eqref{Eq:phi2_F}}, which comes from the constraint
$\bar{M}^{2}_{ \mathrm T}(0)=0$. Both systems of equations contain a
self-consistent self-energy function to be solved for, which we
approximate using iterations.\footnote{In certain cases the convergence
of iterations have to be improved using the under-relaxation method
(see e.g. \cite{Marko:2012wc}).} We discretize the self-energy over a
certain set of $N_{\mathrm{S}}$ momentum values, use spline
interpolation to have its value for other momenta and compute all
integrals using the GNU Scientific Library \cite{GSL}, similarly as in
\cite{Pilaftsis:2013xna}. However, we use two different
discretizations. A~uniform one, where
%
%eA.1 #&#
\begin{eqnarray}
K_{i} = K_{\mathrm{min}} + i\frac{K_{\mathrm{max}}-K_{\mathrm
{min}}}{N_{\mathrm{S}}-1},
\end{eqnarray}
and a cubic, non-equidistant one which has resolution concentrated in
the IR region. There
%
%eA.2 #&#
\begin{eqnarray}
K_{i} = K_{\mathrm{min}} +
i^{3}\frac{K_{\mathrm{max}}-K_{\mathrm{min}}}{(N_{\mathrm{S}}-1)^{3}},
\end{eqnarray}
with $i=0,\dots,N_{\mathrm{S-1}}$ in both cases. $K_{\mathrm{min}}$ and
$K_{\mathrm{max}}$ are, respectively, the smallest and largest
available momentum on the grid. In case of the IR regularization
{\eqref{Eq:IR_reg_mass}} $K_{\mathrm{min}}=0$, while for the IR
regularization {\eqref{Eq:IR_reg_mom}}
$K_{\mathrm{min}}=\kappa$. The discretization itself affects certain
aspects of the IR/UV regularization. First of all, due to the finite
number of available points, even the use of the smooth UV regulator
{\eqref{eq:smoothCO}} requires the use of a sharp cutoff. The
smooth regulator function $R_{\text{smooth}}(K/\Lambda)$ of the form
{\eqref{eq:smoothCO}} has as a function of the momentum $K$ an
inflection at the value of the physical cutoff $\Lambda$ used, that is
at $K=\Lambda$. The value $\sigma=50$ guaranties a sharp variation of
the regulator function around the inflection point and a fast
diminishing of its value as the momentum is increased. We use
$K_{\mathrm {max}}\approx 1.5-2\ \Lambda$, for which the value of the
regulator function is approximately $10^{-13}$. The second implication
of the discretization on the regularization only appears in case of the
IR regularization scheme {\eqref{Eq:IR_reg_mom}}. In this case
the momentum discretization starts at $\kappa$, and therefore any lower
boundary of integration has to be bounded from below by $\kappa$. In
this sense the discussed IR regularization scheme is twofold, as no
integral has zero as a lower boundary.

%f7 #&#
\begin{figure}
\centerline{\includegraphics[width=0.6\textwidth]{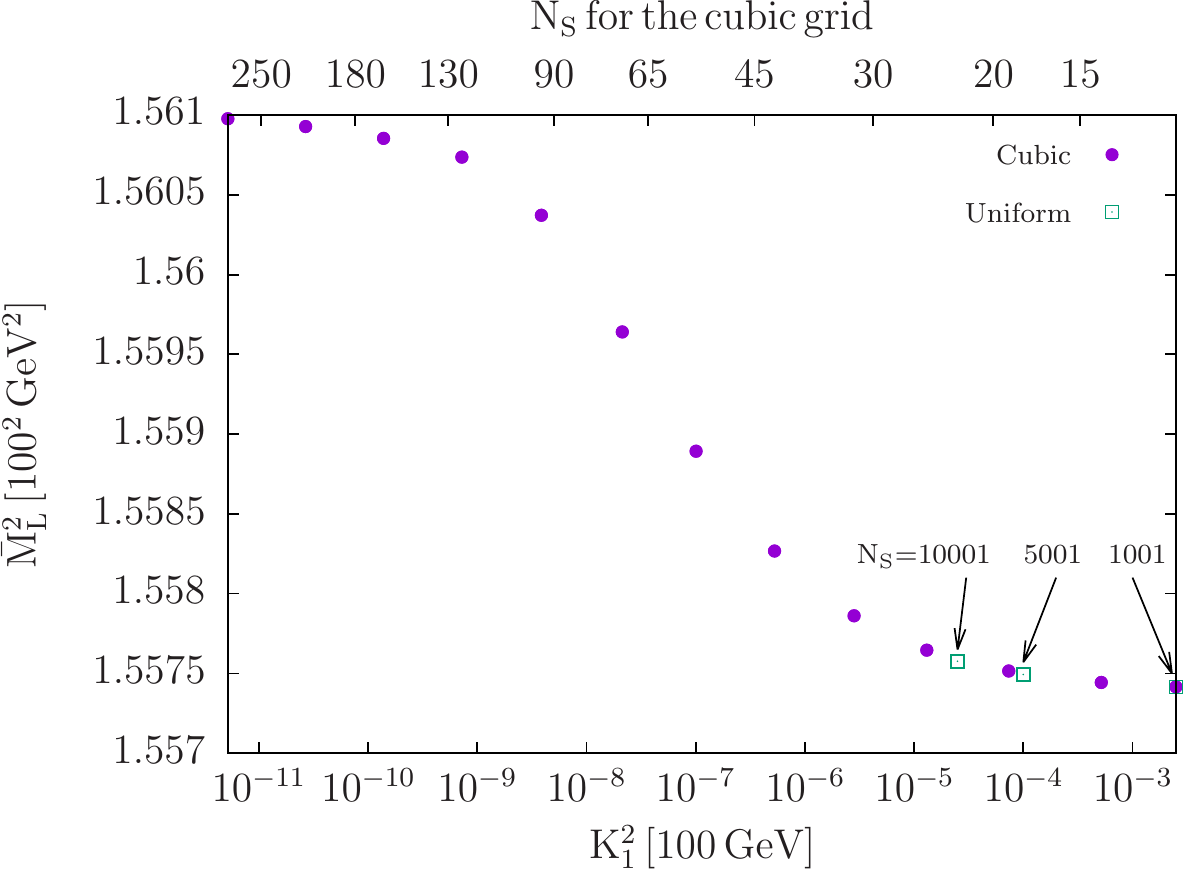}}
\caption{Discretization dependence of the IR regularized $\bar{M}
_{\mathrm{L}}^{2}(K=0)$, where $K_{1}$ is defined as the lowest non-zero
momentum bin in both cubic and uniform grids. We used a sharp UV regulator with
cutoff $\Lambda=5~\mbox{TeV}$ and the IR regularization scheme
$\bar{M}_{\mathrm{T}}^{2}(0)=\kappa^{2}$ with $\kappa=10^{-5}\times 100~\mbox{GeV}$.}
\label{fig:k1}
\end{figure}

The introduction of the cubic grid proves to be very useful, because
high IR resolution can be achieved with a uniform grid only for huge
$N_{\mathrm{S}}$ values. This is illustrated in
{Fig.~\ref{fig:k1}}, where one can see that in comparison to
the uniform grid, the cubic grid needs only a few points and therefore
makes computations more economic.

%sB #&#
\section{Counterterms}\label{app:c-terms}

We describe bellow how to derive with the method summarized in
Sec.~\ref{sec:ren} the explicit relation between bare and renormalized
parameters, that is how to obtain the expression of the counterterms.

First, we subtract the finite gap equations from the bare ones, using
for the tadpole and bubble integrals the decomposition into finite and
divergent pieces. The latter are given in {\eqref{Eq:div_pieces}} and we
use for the local gap mass squared $\bar{M}^{2}_{\text{L/T,l}}$ their
explicit expression which can be read off from {\eqref{Eq:finite_gap}}.
We end up with two equations, one coming from the transverse sector and
one from the longitudinal one, in which the finite quantities
$\phi^{2}$ and $\mathcal{T}_{\text{F}}[\bar{G}_{\text{L/T}}]$ are
multiplied by a combination of bare couplings and divergent integrals.
The bare couplings (or counterterms) are obtained by requiring the
vanishing of the coefficients of the above finite quantities,\footnote{There
are 6 conditions (3 from each equation) for 4 bare
couplings, but it turns out that, as consistency requires, the
conditions determining $\lambda_{0}^{(A)}$ and $\lambda_{0}^{(B)}$ in
the transverse sector are equivalent with the conditions in the
longitudinal sector.} while the remnant gives the unique relation
%
%eB.1 #&#
\begin{eqnarray}
m_{0}^{2}=m^{2}-\frac{\lambda_{0}^{(A+B)}}{6}\big[{\mathcal{T}}_{T=0}[G_{
\mu}]+(\mu^{2}-m^{2}){\mathcal B}_{T=0}[G_{\mu}]\big]\,.
\label{Eq:m2_0}
\end{eqnarray}
From the vanishing of the coefficients of $\mathcal{T}_{\text{F}}[
\bar{G}_{\text{L}}]$ and $\mathcal{T}_{\text{F}}[\bar{G}_{\text{T}}]$ we
obtain the relations
%
%eB.2 #&#
\begin{eqnarray}
\frac{1}{\lambda_{0}^{(A+B)}}=\frac{1}{2\lambda}-\frac{{\mathcal B}_{T=0}[G_{
\mu}]}{6},\quad\frac{1}{\lambda_{0}^{(B)}}=\frac{1}{\lambda}-\frac
{{\mathcal B}_{T=0}[G_{
\mu}]}{6}.
\label{Eq:l0_A_B}
\end{eqnarray}
The vanishing of the coefficient of $\phi^{2}$ in the two equations
determines $\lambda_{2}^{(A/B)}=\lambda+\delta\lambda_{2}^{(A/B)}$.
We split the counterterms into a local and a nonlocal part and write
$\lambda_{2}^{(A/B)}=\lambda_{\text{2,l}}^{(A/B)}+\delta
\lambda_{\text{2,nl}}^{(A/B)}$, where the nonlocal part given by
$\delta\lambda_{\text{2,nl}}^{(B)}=2\delta\lambda_{\text
{2,nl}}^{(A)}=\frac{2
\lambda^{2}}{3}{\mathcal B}_{T=0}[G_{\mu}]$ is used to remove the
divergence of the bubble integral in the self-energy. The two conditions
determining the local part of the counterterms can be written in a
compact form using the pairs $\alpha=A+2B$, $\beta=A$ and
$\alpha=A$, $\beta=A+2B$ as follows:
%
%eB.3 #&#
\begin{eqnarray}
\delta\lambda_{\text{2,l}}^{(\alpha)}&=&\frac{\lambda\lambda
_{0}^{(\alpha)}}{4}\left[ {\mathcal B}_{T=0}[G_{
\mu}]-\frac{5\lambda}{9}{\mathcal{T}}_{\mathrm{d}}^{({\mathcal B})}\right]
%\nonumber\\
%&&
+\frac{\lambda\lambda_{0}^{(\beta)}}{12}\left[ {\mathcal B}_{T=0}[G_{
\mu}]-\frac{\lambda}{3}{\mathcal{T}}_{\mathrm{d}}^{({\mathcal B})}\right] ,
\label{Eq:l2l_A_B}
\end{eqnarray}
with
%
%eB.4 #&#
\begin{eqnarray}
{\mathcal{T}}_{\mathrm{d}}^{({\mathcal B})}\equiv\int_{Q}^{T=0}
G_{\mu}^{2}(Q)\,\Big[{\mathcal B}_{T=0}[G_{
\mu}](Q)-{\mathcal B}_{T=0}[G_{\mu}]\Big]\,.
\end{eqnarray}
Note that the expressions {\eqref{Eq:m2_0}},
{\eqref{Eq:l0_A_B}}, and {\eqref{Eq:l2l_A_B}} are
identical in form with those coming in \cite{Marko:2013lxa} from
renormalization prescriptions imposed at temperature $T_{\star}$.

Now, as a last step, one can check that, by using the relations between
bare and finite parameters, the finite gap equations are obtained from
the bare ones. For this one first uses {\eqref{Eq:m2_0}} and
{\eqref{Eq:l2l_A_B}} in the bare gap equations. Then, one
constructs two combinations
$\bar{M}^{2}_{\text{L}}(K)-\bar{M}^{2}_{\text{T}}(K)$ and
$\bar{M}^{2}_{\text{L}}(K)+\bar{M}^{2}_{\text{T}}(K)$ and
applies {\eqref{Eq:l0_A_B}}. After recognizing in
these equations the appearance of the finite tadpoles given in
{\eqref{Eq:finite_Tad}}, one just turns back from the equations
for the two linear combinations to those for $\bar{M}^{2}_{\text{L}}$
and $\bar{M}^{2}_{\text{T}}$, to obtain
{\eqref{Eq:finite_gap}}.

%sC #&#
\section{Perturbative bubble with a sharp regulator}\label{app:sharp}

In this section and for the sake of clarity, $K$ denotes the norm of
$K$ when this is obvious. The perturbative bubble with a sharp 4d
regulator at the level of the propagators has been computed in
\cite{Reinosa:2011cs}. In the case where one of the propagators is
massless, we obtain
%C.1
\begin{align}
\label{eq:pert}
{\mathcal B}_{\mathrm{pert}}[G,G_{0}](K) &= \frac{1}{4\pi^{3}}\int_{0}^{
\Lambda}d Q\,\frac{Q^{3}T(Q,K,0,-1)}{Q^{2}+M^{2}}
\nonumber
\\
& + \frac{1}{4\pi^{3}}\int_{\Lambda-K}^{\Lambda}d Q\, \frac{Q^{3}
\Delta T(Q,K,0,\alpha)}{Q^{2}+M^{2}}\,,
\qquad
\end{align}
where
%C.2
\begin{align}
T(Q,K,0,\alpha)
&=\,\frac{1}{4Q^{2}K^{2}}\left[\vphantom{\sqrt{\frac{1-
\alpha}{1+\alpha}}} 2Q\,K\sqrt{1-\alpha^{2}}+\big(Q^{2}+K
^{2}\big)\arccos\,\alpha\right.
\nonumber
\\
&\left. -\,2|Q^{2}-K^{2}|\arctan\,\left( \frac{Q+K}{|Q-K|}\sqrt{\frac{1-
\alpha}{1+\alpha}}\right) \right] ,
\end{align}
and $\smash{\Delta T(Q,K,0,\alpha)=T(Q,K,0,\alpha)-T(Q,K,0,-1)}$, with
$\alpha\equiv(K^{2}+Q^{2}-\Lambda^{2})/(2K Q)$. In particular
%
%eC.3 #&#
\begin{eqnarray}
T(Q,K,0,-1) & = & \frac{\pi}{2}\frac{1}{\mathrm{Max}(Q^{2},K^{2})}\,.
\end{eqnarray}

To study the small $K$ behavior, we note that, as $K\to0$, we can
always assume that $K<\Lambda$ and thus the first contribution to
{(\ref{eq:pert})} reads
%
%eC.4 #&#
\begin{eqnarray}
\frac{1}{8\pi^{2}K^{2}}\int_{0}^{K} d Q\, \frac
{Q^{3}}{Q^{2}+M^{2}}+\frac{1}{8\pi^{2}}\int_{K}^{
\Lambda} d Q\, \frac{Q}{Q^{2}+M^{2}}\,.
\end{eqnarray}
The integrals can be easily done and the expansion of the result does
not contain any linear term in $K$ at small~$K$.

In the second contribution to {(\ref{eq:pert})}, since
$Q>\Lambda-K$, we can always assume that $Q>K$ in the limit $K\to0$. We
rescale the integration variable by $K$ and shift the new integration
variable by $x=\Lambda/K$. Denoting the integrand by $h(Q,K,L)$, this
yields
\[
\frac{1}{4\pi^{3}}\int_{\Lambda-K}^{\Lambda}d Q\, h(Q,K,\Lambda) =
\frac{1}{4\pi^{3}}\int_{-1}^{0} d \bar{Q}\, h(\bar{Q}+x,1,x).
\]
We can now expand the integrand for $x\to\infty$ and we obtain
%C.5
\begin{align}
\sim\frac{1}{8\pi^{3}}\frac{x^{-1}}{1+\frac{M^{2}}{\Lambda^{2}}}
\int_{-1}^{0}d\bar{Q}\,\left\{ \arccos\,\bar{Q}-\bar{Q}\sqrt{1-
\bar{Q}^{2}}-\pi\right\} .
\end{align}
Computing this last integral, we finally arrive at
%
%eC.6 #&#
\begin{eqnarray}
{\mathcal B}_{\mathrm{pert}}[G,G_{0}](K)-{\mathcal B}_{\mathrm
{pert}}[G,G_{0}]\sim-\frac{1}{12\pi^{3}}\frac{K\Lambda}{\Lambda
^{2}+M^{2}}\,.
\label{eq:pertLinearSmallK}
\end{eqnarray}

In contrast, applying the same considerations to the perturbative bubble
with a sharp regulator applied at the level of the loop, and not at the
level of the propagators, we do not obtain a linear $K$ behavior at
small $K$. Similarly there is no linear behavior in the presence of a
smooth regulator.

At finite temperature, the same bubble with a $3$d sharp propagator
regulator reads
%C.7
\begin{align}
{\mathcal B}_{\mathrm{pert}}[G,G_{0}](K) =&\,
%\nonumber\\
\frac{1}{8\pi^{2}k}\int_{0}^{\Lambda}dq \,q\Bigg[\frac{1+2n_{
\varepsilon_{q}}}{2\varepsilon_{q}}{\mathrm{Re}}\,\ln\frac
{-(\varepsilon
_{q}-i\omega)^{2}+(k+q)^{2}}{-(\varepsilon_{q}-i\omega)^{2}+(k-q)^{2}}
\nonumber
\\
&{}+\frac{1+2n_{q}}{2q}{\mathrm{Re}}\,\ln\frac{-(q-i\omega)^{2}+
\varepsilon_{k+q}^{2}}{-(q-i\omega)^{2}+\varepsilon_{k-q}^{2}}\Bigg]
\nonumber
\\
&{}+\frac{1}{8\pi^{2}k}\int_{\Lambda-k}^{\Lambda}dq \,q\Bigg[\frac{1+2n
_{\varepsilon_{q}}}{2\varepsilon_{q}}{\mathrm{Re}}\,\ln\frac
{-(\varepsilon
_{q}-i\omega)^{2}+\Lambda^{2}}{-(\varepsilon_{q}-i\omega)^{2}+(k+q)^{2}}
\nonumber
\\
&{}+\frac{1+2n_{q}}{2q}{\mathrm{Re}}\,\ln\frac{-(q-i\omega)^{2}+
\varepsilon_{\Lambda}^{2}}{-(q-i\omega)^{2}+\varepsilon_{k+q}^{2}}
\Bigg]\,.
\end{align}
The contributions involving the integrals from $0$ to $\Lambda$ are of
order $k^{2}$ as $k\to0$, as is easily checked by expanding the
integrands. The contributions involving the integrals from $\Lambda-k$
to $\Lambda$ are treated using the change of variables $q=kx+\Lambda
$ and expanding the integrand at small $k$. For instance
%C.8
\begin{align}
&\frac{1}{8\pi^{2}}\int_{-1}^{0} dx \,\left( kx+\Lambda\right) \frac{1+2n
_{\varepsilon_{kx+\Lambda}}}{2\varepsilon_{kx+\Lambda}}
%\nonumber\\ \times
\mathrm{Re}\,\ln\frac{-(\varepsilon_{kx+\Lambda}-i\omega)^{2}+
\Lambda^{2}}{-(\varepsilon_{kx+\Lambda}-i\omega)^{2}+(k+kx+\Lambda
)^{2}}
\nonumber\\
&\quad \sim\frac{k}{8\pi^{2}}\frac{1+2n_{\varepsilon_{\Lambda}}}{
\varepsilon_{\Lambda}}
%\nonumber\\ \times
\frac{\Lambda^{2}(m^{2}-\omega^{2})}{(m^{2}-\omega^{2})^{2}+4
\omega^{2}(\Lambda^{2}+m^{2})}\underbrace{\int_{-1}^{0}dx\,(1+x)}_{1/2}
\,.
\end{align}
Together with a similar contribution from the other integral, we obtain
%
%eC.9 #&#
\begin{eqnarray}
\sim\frac{k}{8\pi^{2}}\Bigg[\frac{1+2n_{\varepsilon_{\Lambda
}}}{2\varepsilon_{
\Lambda}}\frac{\Lambda^{2}(m^{2}-\omega^{2})}{(m^{2}-\omega
^{2})^{2}+4\omega^{2}(\Lambda^{2}+m^{2})}
%\nonumber\\
-\frac{1+2n_{\Lambda}}{2\Lambda}\frac{\Lambda^{2}(m^{2}+\omega
^{2})}{(m^{2}+\omega^{2})^{2}+4\omega^{2}\Lambda^{2}}\Bigg]\,.
\end{eqnarray}
We thus find a linear $k$-dependence at small $k$. Its finite
temperature contribution is suppressed by a factor $e^{-\Lambda/T}$.
The zero temperature contribution is suppressed by a factor
$1/\Lambda$. This is obvious for $\omega\neq0$. For $\omega=0$, one
needs to combine the two contributions
%
%eC.10 #&#
\begin{eqnarray}
\sim\frac{k}{16\pi^{2}}\frac{\Lambda^{2}}{m^{2}}\left[ \frac
{1}{\varepsilon_{
\Lambda}}-\frac{1}{\Lambda}\right] \sim-\frac{1}{32\pi^{2}}\frac
{k}{\Lambda}\,.
\label{eq:finiteT_linearDep}
\end{eqnarray}

\end{document}